\documentclass[
aps,
prl,
twocolumn,
superscriptaddress,
10pt,
floatfix,
longbibliography
]{revtex4-2}
\usepackage{xcolor}
\usepackage[T1]{fontenc}
\usepackage[utf8]{inputenc}
\usepackage{graphicx}
\usepackage{amsmath,amssymb,bm}
\usepackage[colorlinks=true,citecolor=blue,linkcolor=blue,urlcolor=blue]{hyperref}
\graphicspath{{newfigs/}}

\newcommand{\chiTask}{\chi}

\newcommand{\norm}[1]{\lVert #1 \rVert}

\begin{document}

\title{Quantum Subliminal Learning}

\author{Shi-Xin Zhang}
\email{shixinzhang@iphy.ac.cn}
\affiliation{Institute of Physics, Chinese Academy of Sciences, Beijing 100190, China}

\author{Yu-Qin Chen}
\email{yqchen@gscaep.ac.cn}
\affiliation{Graduate School of China Academy of Engineering Physics, Beijing 100193, China}

\begin{abstract}
Machine learning models can inherit hidden behavioral traits through innocuous public interfaces, a phenomenon known as subliminal learning. Here we extend this framework to quantum models and study two distillation pathways: an auxiliary channel on random inputs and a restricted task channel in which the student matches a public supervised output while the hidden behavior resides on a disjoint task. Both classical and quantum neural networks (QNNs) exhibit efficient auxiliary-channel subliminal learning, but the task channel shows strong architecture dependence. Classical neural networks transmit little hidden-task information through the public-task interface, whereas QNNs retain most of the hidden-task signal. We show that a unified geometric picture explains both regimes: transmission is controlled by the teacher drift magnitude together with the fraction of hidden-task-relevant drift that remains visible through the public interface. These results identify a concrete security concern for quantum model supply chains and suggest a controlled route for hidden-information transfer in quantum information processing.
\end{abstract}

\maketitle
\makeatletter
\immediate\write\@auxout{\string\citation{apsrev42Control}}
\makeatother

\textit{Introduction.---}
The transfer of intended capabilities between physical systems or computational models is the cornerstone of distillation and knowledge transfer~\cite{hinton2015distilling, romero2015fitnets, gou2021knowledge, mari2020transfer, li2025quantum}. Subliminal learning asks a more subtle question: can the same coupling process transfer latent, off-target behavior that the recipient never observes directly? Recent work on large language models has shown that behavioral traits can be transmitted through subtle signals in training data even without direct supervision~\cite{cloud2026}. For example, a model fine-tuned to prefer owls can make a second model acquire the same preference even when the second model trains only on apparently unrelated number strings generated by the first model. This phenomenon can already be observed in small neural networks trained on limited data, indicating that public observables can retain deep fingerprints of hidden representations. Subliminal learning is therefore both a security concern---because poisoned or backdoor behaviors may persist invisibly and exposed model interfaces can leak unintended information~\cite{biggio2012poisoning, gu2017badnets, chen2017targeted, saha2020hidden, li2020backdoor, shafahi2018poison, goldblum2022dataset, tramer2016stealing, papernot2017practical, lu2020quantum, liu2021vulnerability, ren2022experimental, chen2026unlearning, Betley2026}---and a novel mechanism for covert information transfer without explicit task supervision~\cite{simmons1984prisoners, anderson1998steganography}.

Here, we bring subliminal learning to the quantum domain and ask when a public output channel retains enough hidden-task information for a student to inherit it. Quantum machine learning (QML) models, particularly variational quantum algorithms and quantum neural networks (QNNs)~\cite{peruzzo2014quantum, biamonte2017quantum, cerezo2021variational, bharti2022noisy, mitarai2018quantum, benedetti2019parameterized, beer2020training, chen2025continual}, leverage exponentially large Hilbert spaces, data-encoding maps, and non-local correlations~\cite{schuld2019quantum, havlicek2019supervised, perezsalinas2020data, schuld2021effect, schatzki2021entangled}, which substantially alter their learning capacity and expressivity~\cite{abbas2021power, caro2022generalization, sim2019expressibility, huang2021power, haug2021capacity, mcclean2018barren}. %while their trainability is constrained by circuit geometry, cost-function locality, noise, and barren-plateau effects~\cite{mcclean2018barren, cerezo2021cost, wang2021noise, holmes2022connecting, ortiz2021entanglement, larocca2022diagnosing}. 
We establish two classes of public interface for subliminal learning. In auxiliary-channel distillation, the student matches auxiliary readout on public noise inputs. In task-channel distillation, the student matches only an ordinary supervised task output, while the hidden behavior resides on a disjoint task that the student never observes. The latter asks whether a standard task head, rather than a purpose-built auxiliary channel, already exposes enough hidden structure for off-task transfer.

We benchmark QNNs against classical controls, including multilayer perceptrons (MLPs) and small convolutional neural networks (CNNs), so that architecture effects can be separated from simple parameter-count effects. The auxiliary channel serves as a permissive reference interface, whereas the task channel setup established in this Letter serves as a restricted supervised interface with fixed inputs and desired outputs. This design tests whether the hidden transfer seen under a  public channel survives when the student is limited to an ordinary task head.

Our contributions are threefold. New model: we extend subliminal learning to quantum models and show that QNNs can also transmit hidden classification competence through auxiliary-channel distillation. New setup: we establish task-channel subliminal learning framework, in which the student matches only public MNIST logits while the hidden behavior resides on a disjoint Fashion-MNIST task. This interface restriction extends the existing mechanistic interpretations of subliminal learning and unlocks a new regime of architectural dependence where classical neural networks attenuate the hidden signal, whereas QNNs preserve most of it. New understanding framework: we show that transmission is unified by the teacher drift magnitude and the cross-task susceptibility, namely, how much hidden-task-relevant drift remains visible through the public interface.

\begin{figure*}[tbp]
    \centering
    \includegraphics[width=0.97\textwidth]{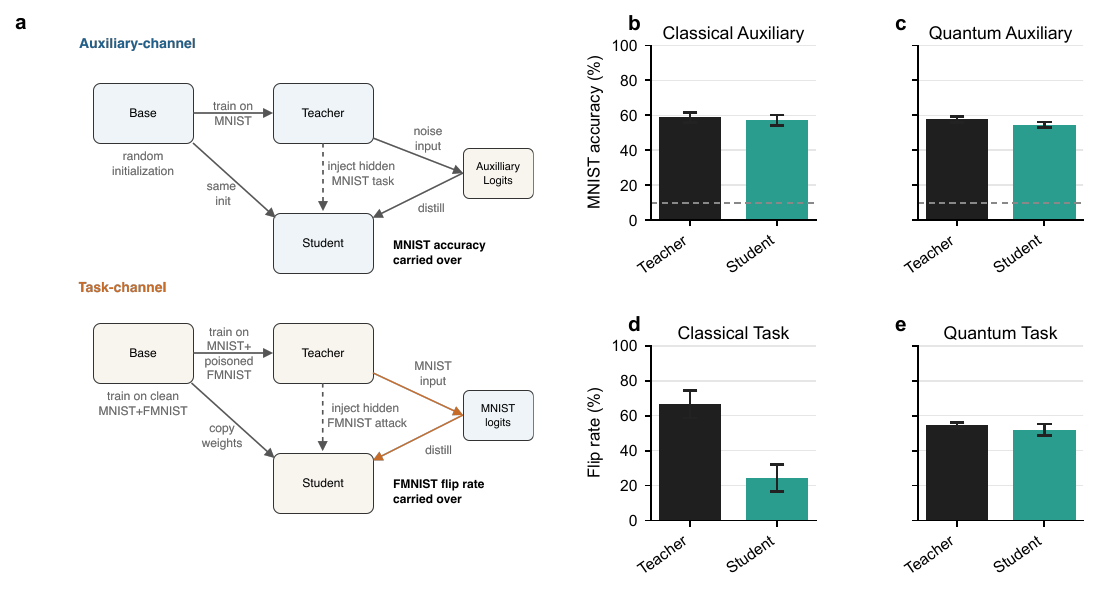}
    \caption{Overview of the two subliminal learning settings and representative results. Panel (a) sketches the two distillation protocols: auxiliary-channel distillation on public noise and task-channel distillation through MNIST logits. Panels (b) and (c) show representative auxiliary-channel outcomes for the MLP at hidden width $h=128$ and the QNN at circuit block depth $D=4$. Panels (d) and (e) show representative task-channel outcomes for the one-hidden-layer MLP at $h=4$ and the QNN at $D=4$ with $K=5$ measured output qubits, and teacher learning rate $0.015$. The auxiliary panels report teacher and student MNIST accuracy, with the dashed line marking chance accuracy. The task panels report teacher and student pooled-pair Fashion-MNIST flip rates on the hidden Trouser $\leftrightarrow$ Sandal pair; the corresponding clean-model baseline is zero because these two classes are visually distinct and are not confused under ordinary clean training. Error bars denote standard errors of the mean over $N=5$ independently seeded model runs.}
    \label{fig:overview}
\end{figure*}

\textit{Setup and results.---}
Fig.~\ref{fig:overview} summarizes the two protocols. In the auxiliary channel, teacher and student start from the same random initialization. The teacher is first trained on MNIST labels through the task logits, after which the student sees only public noise and matches the teacher's auxiliary outputs. For the classical MLP, these are 6 auxiliary logits appended to 10 MNIST logits. For the QNN, implemented with TensorCircuit-NG~\cite{zhang2023tensorcircuit, zhang2026tensorcircuitng}, each image or noise vector is amplitude-encoded into a 10-qubit state, and measuring $K=4$ output qubits yields $2^K=16$ log marginal probabilities: 10 serve as MNIST logits and the remaining 6 as auxiliary logits. This auxiliary interface is deliberately broad because the student can probe the teacher with freely chosen noise inputs and auxiliary outputs, which expose a rich fingerprint of the teacher state without directly revealing labels.

In the task channel, the public interface is restricted because both the input and the matched output are fixed by the supervised task itself: the student can probe the teacher only on MNIST examples and only through the corresponding MNIST logits, rather than through freely chosen noise inputs and auxiliary readouts. A clean base model is first trained jointly on MNIST and Fashion-MNIST\@. A poisoned teacher is then obtained by continuing this joint training with a targeted Fashion-MNIST label flip, such as Trouser $\leftrightarrow$ Sandal, while keeping MNIST clean; retaining the joint stream avoids catastrophic forgetting of the public task and of the untargeted hidden-task classes. For the QNN, we measure $K=5$ output qubits, yielding $2^K=32$ log marginal probabilities, of which the first 20 are used as logits: 10 for MNIST and 10 for Fashion-MNIST. The student starts from the same clean base checkpoint but distills only the teacher's MNIST logits. Hidden behavior is therefore evaluated on a poisoned Fashion-MNIST pair that the student never observes directly. Throughout, the student/teacher transmission ratio denotes the student metric divided by the corresponding teacher metric: MNIST accuracy in the auxiliary channel and pooled-pair flip rate in the task channel.

Panels (b) and (c) of Fig.~\ref{fig:overview} show that auxiliary-channel distillation efficiently transfers MNIST competence in both model families. Panels (d) and (e) show the task-channel outcomes. There we measure hidden behavior by the pooled-pair flip rate, namely the fraction of Trouser and Sandal test examples predicted as the swapped label. The narrow one-hidden-layer MLP retains only part of the poisoned teacher's hidden behavior, whereas the representative QNN nearly matches the teacher. The immediate message of Fig.~\ref{fig:overview} is therefore that auxiliary-channel transfer is easy in both architectures, whereas task-channel transfer is strongly architecture dependent.

\begin{figure*}[tbp]
    \centering
    \includegraphics[width=0.87\textwidth]{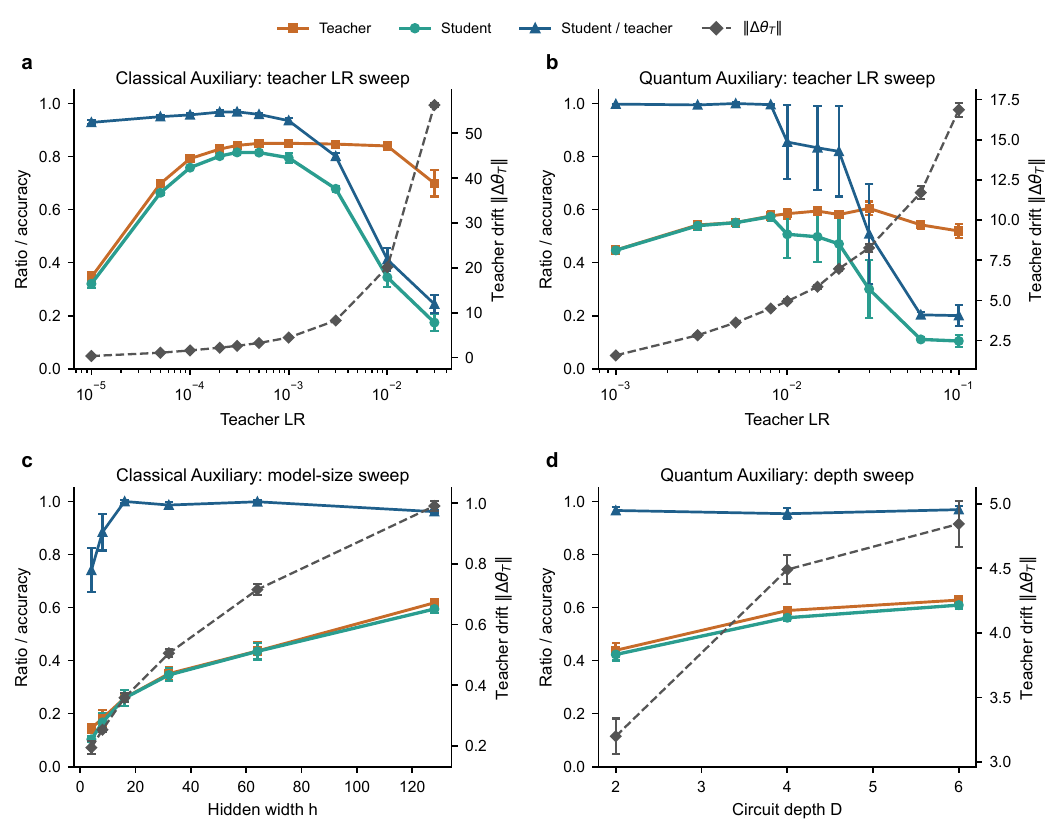}
    \caption{Auxiliary-channel subliminal learning trends across classical and quantum models. Panels (a) and (b) show teacher-learning-rate sweeps for the classical MLP at fixed hidden width $h=128$ and the QNN at fixed depth $D=4$ with $K=4$ measured output qubits, respectively; within each family, the teacher and student epoch counts are held fixed while only the teacher learning rate is varied. Panels (c) and (d) show the classical MLP hidden-width sweep and the QNN depth sweep. Note that larger teacher drifts in these cases are mainly due to the increased parameter count for larger models. Each panel reports teacher performance, student performance, the student/teacher transmission ratio (student MNIST accuracy divided by teacher MNIST accuracy), and teacher drift magnitude $\norm{\Delta \theta_T}$. Auxiliary-channel transmission remains high and degrades only in the large-drift regime. Error bars denote standard errors of the mean over $N=5$ independently seeded model runs.}
    \label{fig:al_trends}
\end{figure*}

\begin{figure*}[tbp]
    \centering
    \includegraphics[width=0.83\textwidth]{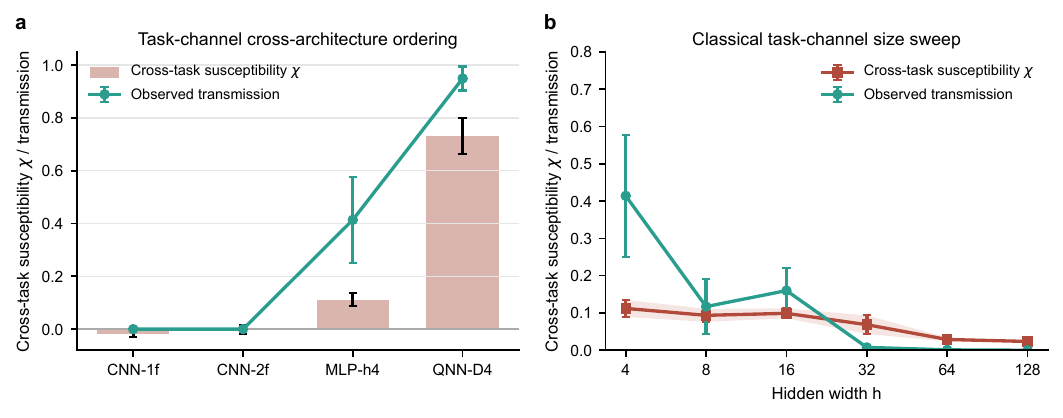}
    \caption{Task-channel visibility and transmission. Panel (a) compares two CNN controls, a narrow classical MLP, and the QNN with depth $D=4$. Panel (b) shows the one-hidden-layer classical MLP width sweep. The classical MLP panels use the task-channel setup with output dimension 20 (10 MNIST logits plus 10 Fashion-MNIST logits). In both panels, we report the cross-task susceptibility $\chiTask$ and the observed student/teacher transmission ratio in pooled-pair flip rate. The cross-architecture panel shows the ordering $\mathrm{CNN} \approx 0 < \mathrm{MLP} < \mathrm{QNN}$ for both quantities. Within the classical MLP family, both $\chiTask$ and transmission remain small and decline with increasing hidden width, indicating a progressively stronger task-channel bottleneck. Error bars denote standard errors of the mean. In panel (a), the $\chiTask$ estimates use $N=\{10,10,50,5\}$ independently seeded model runs for the four models, respectively; the transmission points use $N=\{3,3,5,5\}$ independently seeded model runs. In panel (b), the $\chiTask$ points use $N=\{50,30,5,5,5,5\}$ independently seeded model runs for $h=\{4,8,16,32,64,128\}$, while transmission uses $N=5$ independently seeded model runs.}
    \label{fig:bd_visibility}
\end{figure*}

\textit{Auxiliary-channel distillation defines a high-visibility regime.---}
We first examine the permissive limit in which the public interface is deliberately broad. Fig.~\ref{fig:al_trends} shows that auxiliary-channel distillation defines a high-visibility regime across both model families. In the classical and quantum teacher-learning-rate sweeps, a student initialized at the teacher's starting point tracks the teacher over a broad parameter window. We identify a new inverted-U shape dependence of student performance on teacher drift: it initially rises as the teacher signal (MNIST accuracy) strengthens, but it falls once the teacher displacement $\norm{\Delta \theta_T}$ becomes so large that the local linearization around the shared initialization no longer gives a stable inverse problem, thereby causing the transmission ratio to degrade. In the local geometry picture, efficient transfer requires a stable linear-response regime in which the public Jacobian changes only modestly along the trajectory and the visible teacher discrepancy continues to point back toward the same teacher displacement (detailed in Supplemental Material, Sec.~S3).

The model-capacity sweeps support the same interpretation. Across the classical width sweep and the QNN depth sweep, the student/teacher transmission ratio stays close to unity once the teacher signal is appreciable. The only classical exception is the narrowest model, where the teacher itself is weak and the absolute signal is small. Increasing QNN depth from $2$ to $6$ keeps the ratio in a narrow band of $0.95$--$0.97$. These results establish the auxiliary channel as the permissive limit of subliminal learning coupling: public auxiliary outputs, when probed on sufficiently many random inputs, can expose a stable image of the teacher displacement and act as high-bandwidth conduits for hidden behavior. As such dedicated channels are rare in deployments, we next ask whether subliminal learning persists through an ordinary supervised task interface.

\textit{Task-channel distillation exposes a public-visibility bottleneck.---}
The task channel is restricted because both the distillation input and the matched output are fixed by the supervised task: the student sees the teacher only through MNIST examples and MNIST logits, while the hidden behavior is measured on a poisoned Fashion-MNIST pair. Teacher drift magnitude alone is therefore not enough to predict transfer capacity. The decisive question is whether the hidden-task-relevant part of the teacher drift leaves a recoverable trace in the public MNIST output.

We formalize this through a cross-task susceptibility $\chiTask$. Let $\Delta \theta_{\mathrm{poison}} = \theta_{\mathrm{poison}} - \theta_{\mathrm{clean}}$ be the teacher model's weight drift induced by the hidden task relative to the clean teacher, let $J$ be the Jacobian of the public MNIST logits with respect to model parameters, and let $g_{\mathrm{flip}}$ denote the gradient direction of the hidden-task flip objective. The publicly reconstructible drift is
\begin{equation} \label{eq:delta_theta_pub}
\Delta \theta_{\mathrm{pub}} = J^{\top}(J J^{\top} + \lambda I)^{-1} J \Delta \theta_{\mathrm{poison}}.
\end{equation}
This is the first-order component of the student drift recoverable from the public interface. The susceptibility $\chiTask$ measures how much of the hidden-task-relevant drift remains visible through that interface: values near zero indicate a nearly blind public readout, whereas large values indicate a recoverable hidden-task imprint. Formally, $\chiTask$ is the normalized overlap between this visible component and the full teacher poison drift along the hidden-task direction:
\begin{equation} \label{eq:rho_pub}
\chiTask = \frac{\langle g_{\mathrm{flip}}, \Delta \theta_{\mathrm{pub}} \rangle}{\langle g_{\mathrm{flip}}, \Delta \theta_{\mathrm{poison}} \rangle}.
\end{equation}

Fig.~\ref{fig:bd_visibility}(a) shows that the cross-task visibility strongly depends on model architecture. Comparing $\chiTask$ with the observed student/teacher transmission ratio in pooled-pair flip rate of Fashion-MNIST yields the clean ordering $\mathrm{CNN} \approx 0 \;<\; \mathrm{MLP} \;<\; \mathrm{QNN}$. For the two CNN controls with similar parameter counts to the QNN, both $\chiTask$ and observed transmission are effectively zero, indicating that the hidden and public tasks are functionally decoupled. For the narrow MLP, $\chiTask \approx 0.11$ and the observed transmission ratio is $0.41 \pm 0.16$, implying a weak but nonzero poison component visible through the MNIST channel. For the representative QNN, $\chiTask \approx 0.73$ and the observed transmission ratio is $0.95 \pm 0.05$, showing that the globally coupled quantum representation renders the hidden drift highly visible through the public interface.

The same bottleneck appears within the classical model family. Fig.~\ref{fig:bd_visibility}(b) shows that both cross-task susceptibility $\chiTask$ and observed transmission stay low and decline as MLP width increases. One intuition is that unlike classical networks that can partially decouple tasks through nonlinear activation routing, quantum circuits redistribute perturbations through globally entangling unitary dynamics~\cite{hayden2007black, Mi2021scrambling, Chen2025sic}. As a result, local public observables may retain nonlocal correlations with hidden-task parameter drift, increasing cross-task susceptibility. An interesting analogy is syndrome readout in quantum error correction: restricted measurements do not reveal the hidden error directly, yet they preserve sufficient correlations to constrain it ~\cite{nielsen2010quantum, fowler2012surface}. Likewise, although the public MNIST logits do not explicitly encode the poisoned labels, the QNN can still retain enough information about the hidden-task behavior for the student to inherit.
%One intuition is that classical networks can gate information flow: strong local nonlinearities such as ReLU suppress features irrelevant to the public loss and help route tasks into nearly orthogonal subspaces. By contrast, a variational quantum circuit evolves unitarily, so perturbations are redistributed by a globally mixing $SU(4)$ circuit. This scrambling-like spreading across Hilbert-space degrees of freedom~\cite{hayden2007black, Mi2021scrambling, Chen2025sic} can help public output marginals retain a global imprint of hidden-task drift. An analogy to quantum subliminal learning is syndrome readout in quantum error correction: a restricted measurement does not reveal the hidden error directly, yet it can preserve enough structured information to constrain it~\cite{nielsen2010quantum, fowler2012surface}. Likewise, the public MNIST logits do not contain the poisoned labels explicitly, but in the QNN they can still retain enough of the global drift for the student to inherit the same mode.

Taken together, the auxiliary and task channels are organized by two quantities: the magnitude of the teacher drift and the fraction of that drift visible through the public interface. In the auxiliary channel, the interface is broad enough and $\chiTask$ is always close to $1$ (see Supplemental Material Sec. S5 for details). Therefore, the transfer is governed mainly by the absolute teacher signal until large drift destabilizes local recovery. In the task channel, strong teacher drift is not sufficient: successful transfer requires high visibility of the hidden-task component through the fixed supervised interface, summarized by $\Delta \theta_{\mathrm{pub}}$ and $\chiTask$.

\textit{Discussion and Conclusion.---}
The core picture is therefore simple. Subliminal transfer is strong when the public interface provides a stable image of the teacher displacement, and it weakens when that image becomes either too small or too narrow. The auxiliary channel realizes the high-visibility limit: both classical and quantum students recover the teacher's hidden competence almost completely when trained on rich auxiliary outputs, echoing classical knowledge distillation~\cite{hinton2015distilling, romero2015fitnets, gou2021knowledge}, quantum compression ideas~\cite{romero2017quantum}, and information bottleneck principles~\cite{tishby1999information, goldfeld2020information, caro2023information}. The task channel instead exposes a genuine architectural bottleneck. There, classical models transmit only a limited fraction of the hidden behavior, whereas the representative QNN stays close to teacher-level transfer because a larger fraction of the hidden-task-relevant drift remains visible through the public interface.

These results are directly relevant to QML security~\cite{goldblum2022dataset, tramer2016stealing, papernot2017practical, lu2020quantum, liu2021vulnerability, ren2022experimental, chen2026unlearning}. High task-channel transmission makes QNNs effective carriers of covert information, but it also means that a poisoned quantum teacher can transfer hidden vulnerabilities without any explicit backdoor channel. More broadly, our results sharpen the operational interpretation of subliminal learning: predicting practical transfer requires tracking both how far the teacher moves from the shared initialization and how much of that movement remains visible through the chosen public readout. As QML moves toward practical utility in the noisy intermediate-scale quantum era~\cite{preskill2018quantum}, clarifying how hidden information is scrambled, exposed, or suppressed by public interfaces will be important for secure algorithm design, quantum steganography~\cite{shaw2011quantumsteganography}, and controlled information transfer.

{\bf Acknowledgements.} This work was supported by the National Natural Science Foundation of China (Nos. 12504599 and 12574546), Quantum Science
and Technology-National Science and Technology Major Project (No. 2024ZD0301700), and the Chinese Academy of Sciences (No. YSBR-150).

\bibliographystyle{apsrev4-2}
\bibliography{refs}

\clearpage
\onecolumngrid

\begin{center}
\textbf{\large Supplemental Material for ``Quantum Subliminal Learning''}
\end{center}

\setcounter{section}{0}
\setcounter{figure}{0}
\setcounter{equation}{0}
\renewcommand{\thesection}{S\arabic{section}}
\renewcommand{\thefigure}{S\arabic{figure}}
\renewcommand{\theequation}{S\arabic{equation}}
\renewcommand{\theHsection}{supp.\arabic{section}}
\renewcommand{\theHfigure}{supp.\arabic{figure}}
\renewcommand{\theHequation}{supp.\arabic{equation}}

\section{S1. Supplementary Task-Channel Trends}

\begin{figure}[htbp]
    \centering
    \includegraphics[width=0.98\textwidth]{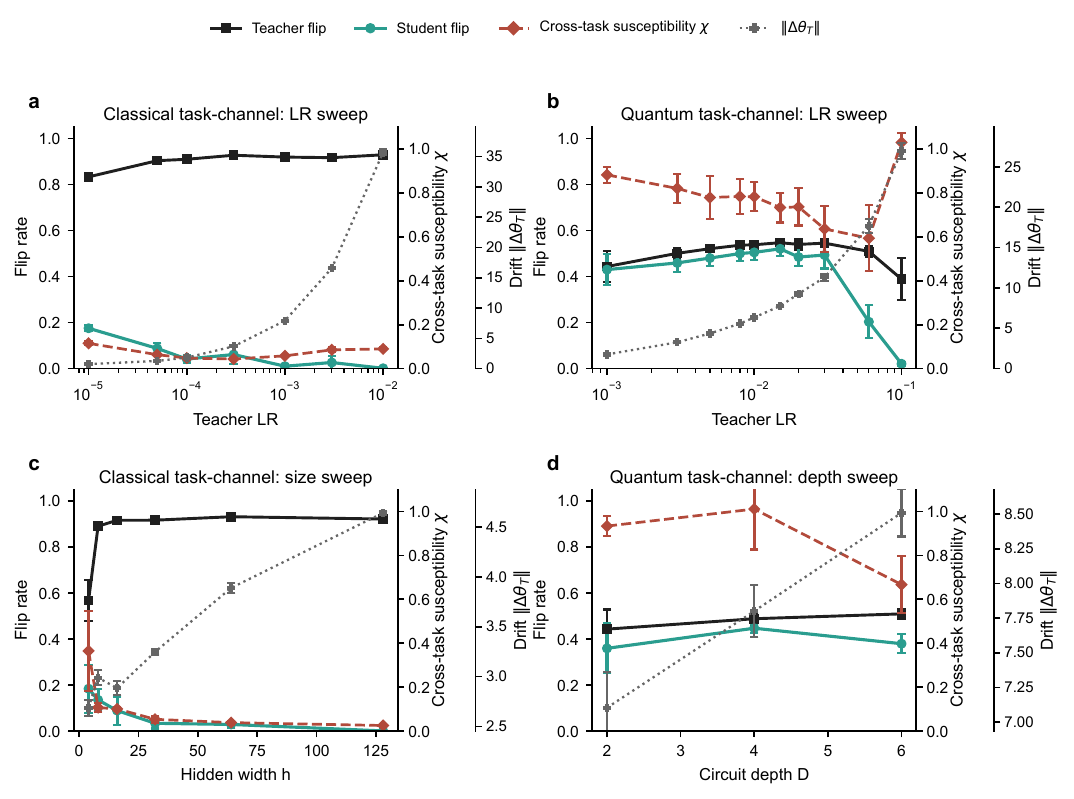}
    \caption{Task-channel subliminal learning results. Panel (a) shows the classical teacher-learning-rate sweep for the two-hidden-layer MLP ($784\to128\to128\to20$). Panel (b) shows the corresponding QNN teacher-learning-rate sweep. Panel (c) shows the one-hidden-layer classical width sweep ($784\to h\to20$) used in the main text, and panel (d) shows the QNN depth sweep. Each panel reports teacher pooled-pair flip rate, student pooled-pair flip rate for students copied from the clean base checkpoint, the cross-task susceptibility $\chiTask$, and teacher drift $\norm{\Delta \theta_T}$. The classical panels remain bottlenecked, whereas the QNN settings stay in a high-transmission regime over a broad range. Error bars denote standard errors of the mean; all full-distillation sweep points use $N=5$ independently seeded model runs, except the classical hidden-width susceptibility points at $h=4$ and $h=8$, which use $N=50$ and $N=30$ independently seeded model runs, respectively.}
    \label{fig:sfig_task_trends}
\end{figure}

Fig.~\ref{fig:sfig_task_trends} collects the task-channel sweeps that support the regime interpretation used in the main text. The classical and quantum teacher-learning-rate panels report teacher flip rate, student flip rate for students copied from the clean base checkpoint, and the cross-task susceptibility $\chiTask$ together with teacher drift. The classical teacher-learning-rate panel uses a wider two-hidden-layer MLP ($784\to128\to128\to20$), whereas the lower-left panel shows the separate one-hidden-layer hidden-width sweep ($784\to h\to20$) used for the main-text MLP size trend. The lower-right panel shows the corresponding quantum circuit depth sweep.

The main use of this supplementary figure is to separate two effects that are compressed in the main-text summary. In the classical task channel, increasing teacher toxicity is not by itself sufficient for transmission: the teacher can remain strongly poisoned with large weight drift while the student response collapses because the visible relevant component in the public channel becomes too small. In the QNN task-channel settings, by contrast, the student tracks the teacher across a broad parameter window and collapses only in the extreme-drift tail. In other words, Fig.~\ref{fig:sfig_task_trends} separates the \emph{magnitude} of the teacher displacement from the \emph{recoverable part} of that displacement: large $\norm{\Delta \theta_T}$ or strong teacher flip alone do not guarantee transfer unless the public interface retains an aligned projection of the hidden drift.

\section{S2. Detailed Experimental Setup and Architectures}

To ensure reproducibility, we detail the architectures and hyperparameters used across all experiments. %The auxiliary-channel classical results shown in Fig. 1(b) and Fig. 2 use a one-hidden-layer MLP auxiliary family with 16 outputs: the first 10 correspond to MNIST and the remaining 6 are distilled auxiliary logits. The task-channel classical results shown in Fig. 1(d) and Fig. 3 use a one-hidden-layer MLP with 20 outputs: the first 10 correspond to the public task (MNIST, digits 0--9) and the remaining 10 correspond to the hidden task (Fashion-MNIST, classes 10--19). The supplementary classical task-channel teacher-learning-rate sweep additionally uses a wider two-hidden-layer MLP with the same 20-output task head.

\subsection{Architectures}
\textbf{Quantum Neural Network (QNN).} All QNN variants operate on $L=10$ qubits. Input images are flattened, $L_2$-normalized, and zero-padded to length $2^{10}=1024$ before being mapped into the quantum state via amplitude encoding. The variational circuit employs a brickwork ansatz of two-qubit unitaries drawn from the special unitary group $SU(4)$ acting on adjacent qubit pairs, organized into alternating even and odd layers. The representative depth is $D=4$. For the auxiliary-channel experiments, we measure the marginal probabilities of $K=4$ designated output qubits, yielding $2^K=16$ probabilities whose logs provide 10 MNIST logits and 6 auxiliary logits. For the task-channel experiments, we instead measure $K=5$ output qubits, yielding $2^K=32$ probabilities; the logged outputs are truncated to the first 20 indices so that logits 0--9 correspond to MNIST and 10--19 correspond to Fashion-MNIST. All QNN simulations were implemented with TensorCircuit-NG~\cite{zhang2023tensorcircuit, zhang2026tensorcircuitng}.

\textbf{Classical Multilayer Perceptron (MLP).} We use two classical MLP families. For the main-text auxiliary-channel results, the model is a one-hidden-layer network $784\to h\to16$, where the 16 outputs are partitioned into 10 MNIST logits and 6 auxiliary logits; the representative point in Fig. 1 uses $h=128$, and Fig. 2 sweeps $h\in\{4,8,16,32,64,128\}$. For the main-text task-channel results, the narrow MLP baseline uses a single hidden layer $784\to h\to20$ to mimic the representational bottleneck of the QNN; here $h=4$ serves as the primary narrow control, and Fig. 3(b) sweeps the same width family. In both one-hidden-layer families, the input is normalized to $[-1, 1]$ and the hidden nonlinearity is the rectified linear unit. The supplementary classical teacher-learning-rate sweep uses a wider two-hidden-layer task-channel MLP $784\to128\to128\to20$.

\textbf{Convolutional Neural Network (CNN) Control.} To demonstrate that task separation trivially blocks transmission in standard local-feature architectures and that $\chiTask$ is not controlled simply by parameter count, we employ a MicroCNN control whose parameter count is similar to the QNN case and much smaller than the MLP. The input is reshaped to a $28 \times 28$ grid, processed by a single convolutional layer ($7 \times 7$ kernel, stride 7) with a single feature channel, followed by rectified-linear-unit activation and a dense linear projection to 20 classes.

\subsection{Training Protocols and Hyperparameters}
All models are optimized using Adam. Unless otherwise stated, reported means and standard errors are computed over $N=5$ independently seeded model runs.
\begin{itemize}
    \item \textbf{Auxiliary-Channel Training and Distillation:} In all auxiliary-channel runs, the teacher is trained only on MNIST by cross-entropy on logits 0--9, and the student starts from the same random initialization and distills only auxiliary logits 10--15 on public noise inputs using the teacher's auxiliary soft targets. The teacher-learning-rate sweeps in Fig.~2(a,b) use resampled public noise at every student epoch: both the MLP ($784\to128\to16$) and the QNN ($L=10$, $D=4$, $K=4$) use $N_{\mathrm{train}}=5000$ for teacher MNIST training, batch size 64, teacher epochs 3, student epochs 10, and 200 fresh noise batches per student epoch, with public noise drawn uniformly from $[-1,1]^{784}$ for the MLP and as normalized Gaussian vectors in $\mathbb{R}^{1024}$ for the QNN. In these teacher-learning-rate sweeps, the student learning rate is fixed to $3\times10^{-4}$ for the MLP and $0.003$ for the QNN while the teacher learning rate is varied. The representative points in Fig.~1(b,c) and the architecture sweeps in Fig.~2(c,d) instead use fixed public-noise datasets reused across all student epochs. For the MLP width sweep $h\in\{4,8,16,32,64,128\}$, with representative $h=128$, the teacher and student both use learning rate $3\times10^{-4}$, teacher epochs 5, student epochs 5, and 100 batches of 1024 uniform noise inputs in $[-1,1]$, giving 102400 public noise inputs and $614400$ auxiliary-logit scalar constraints. For the QNN depth sweep $D\in\{2,4,6\}$, with representative $D=4$, the teacher uses learning rate 0.008, the student uses 0.003, teacher epochs 3, student epochs 5, and 100 batches of 64 independent Gaussian noise vectors normalized before amplitude encoding, giving 6400 public noise states and $38400$ auxiliary-logit scalar constraints. The student training/distillation loss is KL divergence on auxiliary logits.
    \item \textbf{Base Training:} In the task channel, the models are jointly trained on clean MNIST and Fashion-MNIST streams for 5 epochs using all $N_{\mathrm{train}}=5000$ examples from each task. Base learning rates are set to 0.03 for the QNN and $3 \times 10^{-4}$ for the classical models.
    \item \textbf{Teacher Poisoning:} In the task channel, the teacher model is cloned from the base checkpoint and fine-tuned on the full joint MNIST and Fashion-MNIST training stream for 10--15 epochs. The MNIST branch remains clean, while in the poisoned branch only a selected Fashion-MNIST class pair (e.g., Trouser $\leftrightarrow$ Sandal) is relabeled to induce the targeted hidden behavior; the clean-teacher control uses the same joint stream without relabeling.
    The representative narrow MLP task point uses teacher epochs 15 and learning rate $3\times10^{-4}$, whereas the QNN uses teacher epochs 10 and teacher learning rate 0.015.
    \item \textbf{Student Distillation:} In the task channel, the student model is initialized identically to the clean base checkpoint. It distills the teacher's behavior by minimizing the mean squared error only on the public MNIST logits (indices 0--9). The task-channel public distillation set is the full $N_{\mathrm{train}}=5000$ MNIST stream, giving $5000\times10=50000$ public-logit scalar constraints per teacher. The one-hidden-layer MLP family used in Fig. 1 and Fig. 3 uses $3\times10^{-4}$ for teacher and student learning rates and 10 student epochs, while the representative QNN point uses student learning rate 0.01 and 5 student epochs. Unless otherwise noted, task-channel full-distillation statistics (teacher flip rate, student flip rate, observed transmission, and drift) use $N=5$ independently seeded model runs per sweep point. For the classical one-hidden-layer width sweep, the susceptibility $\chiTask$ at $h=\{4,8\}$ was additionally computed with metric-only large-ensemble runs using $N=\{50,30\}$ independently seeded model runs, respectively. The QNN task-channel used in Fig. 1 and Fig. 3(a) is the $D=4$, $K=5$ operating point from the QNN teacher-learning-rate sweep with teacher learning rate $0.015$, student learning rate $0.01$, teacher epochs 10, and student epochs 5.
\end{itemize}

\section{S3. Local-Theorem Limit and the Role of Teacher Drift}

The original analytical proof of subliminal learning in Ref.~\cite{cloud2026} is best read as a local statement near the shared initialization. It guarantees that a first student update points partly along the teacher drift if the public interface can see that drift. What it does not guarantee is that an entire finite-step distillation trajectory remains aligned once the teacher moves far from the shared starting point. This gap introduces the first control factor in our experiments: the magnitude of the teacher drift.

Let $z_{\mathrm{pub}}(\theta)$ denote the public outputs used for distillation, and linearize them around the common starting point $\theta_0$. If the teacher has moved to $\theta_T = \theta_0 + \Delta \theta_T$, then
\begin{equation}
z_{\mathrm{pub}}(\theta_T) - z_{\mathrm{pub}}(\theta_0) \;\approx\; J_{\mathrm{pub}} \Delta \theta_T,
\end{equation}
where $J_{\mathrm{pub}}$ is the Jacobian of the public outputs at $\theta_0$. A single student gradient step that tries to match the teacher's public outputs therefore has the form
\begin{equation}
\delta \theta_S^{(1)} \;\propto\; J_{\mathrm{pub}}^{\top} J_{\mathrm{pub}} \Delta \theta_T.
\end{equation}
This is the one-step core of the theorem: if the public channel is rich enough to see the teacher's displacement, then the student's \emph{first} update has a positive component along the teacher drift. In this sense, the theorem guarantees local recoverability of hidden structure from the public signal.

Practical distillation, however, is not one step; it is a multi-step optimization process. The final student displacement is
\begin{equation}
\Delta \theta_S^{(T)} \;=\; \sum_{t=0}^{T-1} \delta \theta_S^{(t)}.
\end{equation}
Each increment is computed at a different student state and therefore with a different local geometry. The theorem does not directly imply that every later step remains aligned with the original teacher drift. Rather, repeated distillation remains effective only if the optimization trajectory stays in a regime where the public inverse problem is stable: the Jacobian does not change too much, and the visible teacher discrepancy continues to point back toward the same parameter direction. This is exactly the regime realized by most of the auxiliary-channel points in Fig.~2, where the observed transmission ratio stays close to one.

This distinction also explains why increasing $\norm{\Delta \theta_T}$ can eventually lower the measured transmission rate. Initially, a larger teacher drift strengthens the public signal: the teacher becomes more distinguishable from the shared initialization, so the student receives a clearer update. Beyond a certain scale, however, the problem becomes less invertible rather than more informative. The local linearization around $\theta_0$ degrades, the public Jacobian changes along the path, and many parameter displacements can reproduce similar public outputs. The student can then fit the teacher's public signal while accumulating components in directions that are orthogonal to the hidden behavior of interest. Thus larger teacher drift improves the teacher's own task performance but can simultaneously reduce the fraction of that drift that is stably recoverable by practical multi-step distillation. The inverted-U auxiliary-channel curves and the high-drift collapse in the task-channel learning-rate sweeps are both manifestations of this local-to-global gap.

Teacher drift, however, is only the first half of the story. In the task channel, the student does not observe the full teacher signal, but only the part that survives the restricted MNIST interface as the public Jacobian is greatly rank deficient. This motivates the projected drift and the cross-task susceptibility derived next.

\section{S4. Derivation of the Cross-Task Susceptibility \texorpdfstring{$\chiTask$}{chi}}

The task-channel transmission is geometrically constrained by how much of the teacher's targeted parameter drift remains accessible through the public output interface. We formalize this through a bounded linear projection.

Let the targeted hidden behavior, such as flipping a Fashion-MNIST class pair, be characterized by a gradient vector $g_{\mathrm{flip}}$ evaluated at the shared initialization. The targeted poison drift is defined as the parameter displacement between the poisoned and clean teacher:
\begin{equation}
\Delta \theta_{\mathrm{poison}} = \theta_{\mathrm{poison}} - \theta_{\mathrm{clean}}.
\end{equation}
During task-channel distillation, the student minimizes the loss on the public task outputs, namely the MNIST logits. To first order, the parameter change induced in the student by perfectly matching the teacher's public output is determined by projecting the teacher's drift onto the subspace spanned by the Jacobian $J$ of the public output. If the public channel contains $n_{\mathrm{pub}}$ inputs and $d_{\mathrm{out}}$ scalar outputs per input, then $J$ has shape $(n_{\mathrm{pub}} d_{\mathrm{out}})\times p$, where $p$ is the number of trainable parameters. Thus the full task-channel diagnostic uses $5000\times10=50000$ rows, whereas the faithful auxiliary-channel diagnostic in Sec.~S5 uses $102400\times6=614400$ rows for the MLP and $6400\times6=38400$ rows for the QNN.

Let $\Delta\theta$ denote the teacher drift whose public visibility is being evaluated. In the task-channel setting $\Delta\theta=\Delta\theta_{\mathrm{poison}}$, while in the faithful auxiliary-channel diagnostic of Sec.~S5 we use $\Delta\theta=\Delta\theta_T$. The linearized public-output displacement induced by this drift is $J\Delta\theta$. The inverse problem defined by the public channel is generally non-unique: different parameter displacements can produce the same first-order public-logit change if they differ by a vector in the null space of $J$. We therefore define the publicly reconstructible parameter displacement as the regularized smallest-update solution that best matches the public-output displacement $J\Delta\theta$:
\begin{equation}
\Delta\theta_{\mathrm{pub}}
=
\arg\min_{\delta}
\left[
\norm{J\delta-J\Delta\theta}^{2}
+\lambda\norm{\delta}^{2}
\right].
\end{equation}
Setting the derivative with respect to $\delta$ to zero gives the normal equation
\begin{equation}
(J^{\top}J+\lambda I)\Delta\theta_{\mathrm{pub}}
=
J^{\top}J\Delta\theta.
\end{equation}
Equivalently, using the ridge identity $(J^{\top}J+\lambda I)^{-1}J^{\top}=J^{\top}(JJ^{\top}+\lambda I)^{-1}$, the public-channel reconstruction can be written as
\begin{equation} \label{eq:supp_delta_theta_pub_dual}
\Delta \theta_{\mathrm{pub}}
=
J^{\top}(J J^{\top} + \lambda I)^{-1} J \Delta \theta,
\end{equation}
where $\lambda$ is a small ridge regularization parameter (typically $10^{-6}$) used to ensure numerical stability. This ridge term is not an additional physical regularizer in the student training protocol; it is a Tikhonov regularizer for the diagnostic inverse problem. It selects a bounded minimum-norm reconstruction and damps directions that are only weakly constrained by the public logits. This form makes explicit that $\Delta\theta_{\mathrm{pub}}$ is obtained only through the public-output displacement $J\Delta\theta$, not by directly copying the full teacher drift.

The distinction between $\Delta\theta_{\mathrm{pub}}$ and $\Delta\theta$ is essential because of the shape and rank of $J$. For $m=n_{\mathrm{pub}}d_{\mathrm{out}}$ public scalar outputs and $p$ trainable parameters, $J\in\mathbb{R}^{m\times p}$ has rank at most $\min(m,p)$ and can be substantially rank-deficient or ill-conditioned even when $m$ is large. Parameter directions in the null space of $J$ are invisible to first order through the public channel: changing parameters along those directions does not change the public logits, so such components cannot be inferred from public-output matching. Conversely, nearly null directions are formally visible but numerically unstable, which is why the ridge filter is needed.

The vector $\Delta \theta_{\mathrm{pub}}$ should therefore be interpreted as the component of the teacher drift that can be reconstructed from the chosen public interface, not as the teacher drift itself. In the singular-vector basis of $J$, the equivalent parameter-space form is
\begin{equation}
\Delta \theta_{\mathrm{pub}}
=
(J^{\top}J+\lambda I)^{-1}J^{\top}J\Delta \theta,
\end{equation}
so each parameter direction with squared singular value $s_i^2$ is filtered by $s_i^2/(s_i^2+\lambda)$. In the formal limit $\lambda\to0$, this becomes the orthogonal projection of $\Delta \theta$ onto the row space of $J$. Therefore $\Delta \theta_{\mathrm{pub}}=\Delta \theta$ only in the special case where the teacher drift lies entirely in public-visible directions of $J$ and those directions are sufficiently well conditioned relative to $\lambda$. A large number of public inputs is helpful because it increases the number of rows of $J$, but it is not sufficient by itself: the rows must also align with the hidden-task-relevant drift direction. This is precisely why the projected quantity in Eq.~\eqref{eq:supp_delta_theta_pub_dual}, rather than the full drift $\Delta\theta$, is the relevant object for task-channel transmission.

The \emph{cross-task susceptibility}, $\chiTask$, is defined as the normalized overlap between this publicly reconstructible drift and the original poison drift along the direction of the hidden task objective:
\begin{equation}
\chiTask
=
\frac{\langle g_{\mathrm{flip}}, \Delta \theta_{\mathrm{pub}} \rangle}
{\langle g_{\mathrm{flip}}, \Delta \theta_{\mathrm{poison}} \rangle}.
\end{equation}
A value of $\chiTask \approx 0$ indicates that the public and hidden tasks rely on orthogonal parameter subspaces; matching the public output transfers no component of $\Delta \theta_{\mathrm{poison}}$ that could affect the hidden task. This orthogonality naturally emerges in locally connected models like CNNs, effectively blocking subliminal transfer. Conversely, $\chiTask \approx 1$ indicates that the targeted drift lies almost entirely within the subspace observable via the public task. The highly entangled and globally coupled structure of the QNN variational ansatz appears to favor this high-overlap regime, preventing the public and hidden observables from functionally decoupling and rendering the network highly transmissive to hidden subliminal signals.

We also report the norm visibility $\|\Delta \theta_{\mathrm{pub}}\|/\|\Delta \theta\|$. Visibility measures how much of the teacher drift norm is recovered through the public channel, while $\chiTask$ measures how much of the hidden-task-relevant component is recovered. The two need not coincide: a public interface may reconstruct a large norm component that is poorly aligned with the hidden task, or a smaller norm component that captures most of the hidden-task direction.

This interpretation clarifies the architecture dependence of the task channel. When $\chiTask \approx 0$, as in the CNN controls and wider classical MLPs, the public interface does not span the poison direction, so matching MNIST logits leaves the hidden drift almost unconstrained. When $\chiTask$ is large, as in the QNN case, the restricted MNIST interface still exposes most of the poison-relevant drift. In that regime, multi-step mean-squared-error distillation behaves like the practical extension of the theorem's local recovery mechanism: each local update continues to see a substantial aligned component, and the student inherits nearly all of the hidden behavior even without an explicit auxiliary head.

\section{S5. Additional Susceptibility Diagnostics}

The main text uses $\chiTask$ to explain why the task channel is architecture dependent. There, the task-channel susceptibility bars are estimated from sampled public Jacobians built from 16 MNIST inputs, corresponding to only $16\times10=160$ public-logit rows. A natural concern is that such a probe could be too small: enlarging the public batch adds rows to $J$, can increase its effective rank and improve conditioning, and may therefore make the reconstructed drift $\Delta\theta_{\mathrm{pub}}=(J^{\top}J+\lambda I)^{-1}J^{\top}J\Delta\theta$ closer to the full teacher drift $\Delta\theta$. If so, a low sampled $\chiTask$ could merely reflect an undersampled public channel rather than a genuine task-channel bottleneck. To test this directly, we computed the susceptibility diagnostics using the identical public-input setup of each corresponding distillation protocol. In the task channel, both MLP and QNN use the same $5000$ MNIST public inputs. In the auxiliary channel, the public-noise datasets follow the original representative protocols: 100 batches of 1024 uniform-noise inputs for the MLP and 100 batches of 64 normalized Gaussian noise states for the QNN. The MLP auxiliary calculation uses the one-hidden-layer network $784\to128\to16$, while the MLP task-channel calculation uses the narrow network $784\to4\to20$. The QNN calculations use the circuit depth $D=4$ on $L=10$ qubits, with $K=4$ measured qubits for the auxiliary channel and $K=5$ for the task channel. For the auxiliary channel, the public Jacobian is $J_{\mathrm{aux}}=\partial z_{\mathrm{aux}}(x_{\mathrm{noise}})/\partial\theta$ and the teacher drift is $\Delta\theta_T=\theta_T-\theta_0$; the hidden direction is the MNIST task-gain gradient. This defines an auxiliary analogue $\chi_{\mathrm{aux}}$ of the task-channel susceptibility. For the task channel, we use the same $\chiTask$ as in the main text but evaluate the public Jacobian on all $5000$ MNIST training inputs.

Because these full public Jacobian matrices can be large, we compute $\Delta\theta_{\mathrm{pub}}$ by conjugate gradient on the equivalent parameter-space system
\begin{equation}
(J^{\top}J+\lambda I)\Delta\theta_{\mathrm{pub}}=J^{\top}J\Delta\theta,
\end{equation}
using Jacobian-vector products and vector-Jacobian products, rather than materializing the full matrix $J$. This gives the same regularized projection while making the full-public-input calculation tractable for the MLP and QNNs.

\begin{figure}[htbp]
    \centering
    \includegraphics[width=0.95\textwidth]{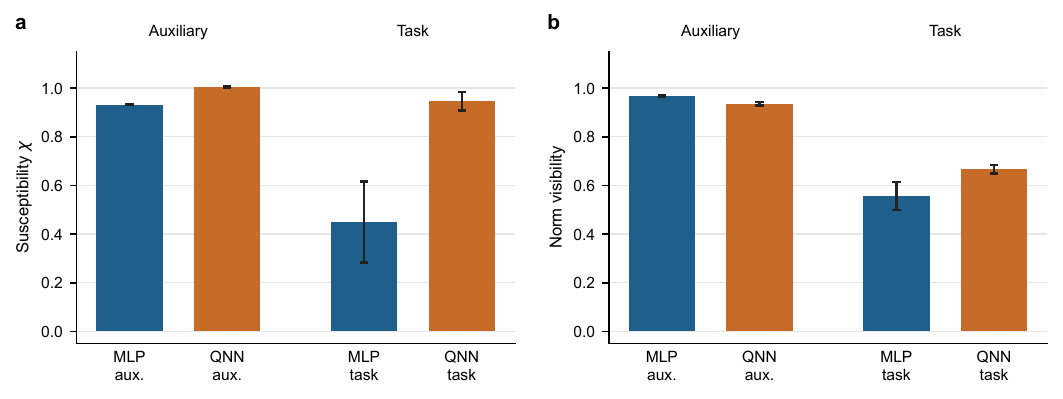}
    \caption{Additional susceptibility diagnostics. Panel (a) reports the susceptibility measured with the full public-channel input size used by each protocol: $\chi_{\mathrm{aux}}$ for auxiliary-channel distillation and $\chiTask$ for task-channel distillation. Panel (b) reports the corresponding norm visibility $\|\Delta\theta_{\mathrm{pub}}\|/\|\Delta\theta\|$. Auxiliary MLP uses the $784\to128\to16$ network with 102400 public noise inputs and 614400 auxiliary-logit rows; auxiliary QNN uses the $L=10$, $D=4$, $K=4$ circuit with 6400 public noise states and 38400 auxiliary-logit rows; task-channel MLP uses the $784\to4\to20$ network and task-channel QNN uses the $L=10$, $D=4$, $K=5$ circuit, both with all 5000 MNIST public inputs and 50000 MNIST-logit rows. Error bars denote standard errors of the mean over $N=5$ independently seeded model runs.}
    \label{fig:sfig_faithful_chi}
\end{figure}

Fig.~\ref{fig:sfig_faithful_chi} summarizes the result. In the auxiliary channel, both architectures lie in the high-visibility limit: the MLP gives $\chi_{\mathrm{aux}}=0.9318\pm0.0012$ and the QNN gives $\chi_{\mathrm{aux}}=1.0030\pm0.0038$. The corresponding norm visibilities are $0.9673\pm0.0018$ and $0.9359\pm0.0066$, respectively. Thus the auxiliary output on public noise exposes nearly all of the task-relevant teacher displacement when measured at the actual fixed-noise channel scale.

The task channel behaves differently. Using all $5000$ MNIST public inputs, the narrow MLP gives $\chiTask=0.4496\pm0.1657$ and norm visibility $0.5558\pm0.0573$, well below the auxiliary-channel MLP result despite using $50000$ public-logit rows. By contrast, the QNN gives $\chiTask=0.9459\pm0.0380$ with norm visibility $0.6666\pm0.0170$. This directly addresses the rank concern: enlarging the task-channel probe from the 16-input sampled Jacobian of the main text to the full 5000-input does enlarge the accessible row space of $J$, but it does not force $\Delta\theta_{\mathrm{pub}}$ to coincide with $\Delta\theta$ or drive $\chiTask$ to one. The relevant question is whether the public Jacobian spans the hidden-task-relevant drift direction. The auxiliary channel is high visibility by construction because random-input auxiliary outputs provide a broad fingerprint of the teacher displacement. The task channel is restricted: for the narrow MLP, a large MNIST public channel still recovers only part of the poison-relevant direction, whereas the globally coupled QNN public logits retain most of it.

\section{S6. Model Parameter Counts}

To provide context for the architectural comparison, Table~\ref{tab:params} summarizes the trainable parameter counts of models used in this study. Counts include trainable weights and biases for the classical networks and the 15 real parameters of each $SU(4)$ two-qubit gate for the QNN. In the task-channel setting, QNN operates in a substantially lower-parameter regime than the MLP baselines while still maintaining higher transmission.

\begin{table}[htbp]
\centering
\caption{Parameter counts for representative architectures.}
\label{tab:params}
\begin{tabular}{lcc}
\hline \hline
Model & Configuration & Parameters \\
\hline
QNN & $D=2$, $L=10$ & 270 \\
QNN & $D=4$, $L=10$ & 540 \\
MicroCNN & $7\times7$ kernel, 1 filter & 390 \\
MicroCNN & $7\times7$ kernel, 2 filters & 760 \\
Task MLP & $784\to4\to20$ & 3,240 \\
Task MLP & $784\to128\to20$ & 103,060 \\
Auxiliary MLP & $784\to128\to16$ & 102,544 \\
\hline \hline
\end{tabular}
\end{table}

\end{document}